\documentclass[preprint,12pt,review]{elsarticle}

\usepackage{lineno,hyperref}
\usepackage{graphicx}
\usepackage{dcolumn}
\usepackage{bm}
\usepackage{float}
\usepackage{epsfig}
\usepackage{booktabs}
\usepackage{subfigure}
\usepackage{graphics}
\usepackage{amssymb}
\usepackage{amsmath}
\usepackage{array}
\usepackage{color}
\usepackage{booktabs}
\usepackage{multirow}
\usepackage{setspace}
\usepackage{nomencl}

\begin{document}

\begin{frontmatter}

\title{A diffuse-interface lattice Boltzmann method for fluid-particle interaction problems}

\author[label1]{Jiao Liu}
\author[label2]{Changsheng Huang}
\author[label1,label3]{Zhenhua Chai\corref{mycorrespondingauthor}}
\cortext[mycorrespondingauthor]{Corresponding author}
\ead{hustczh@hust.edu.cn}
\author[label1,label3]{Baochang Shi}

\address[label1]{School of Mathematics and Statistics, Huazhong University of Science and Technology, Wuhan 430074, China}

\address[label2]{School of Mathematics and Physics, China University of Geosciences, Wuhan 430074, China}

\address[label3]{Hubei Key Laboratory of Engineering Modeling and Scientific Computing, Huazhong University of Science and Technology, Wuhan 430074, China}


\begin{abstract}
In this paper, a diffuse-interface lattice Boltzmann method (DI-LBM) is developed for fluid-particle interaction problems. In this method, the sharp interface between the fluid and solid is replaced by a thin but nonzero thickness transition region named diffuse interface, where the physical variables varies continuously. In order to describe the diffuse interface, we introduce a smooth function, which is similar to the order parameter in phase-field model or the volume fraction of solid phase in the partially saturated lattice Boltzmann method (PS-LBM). In addition, to depict the fluid-particle interaction more accurately, a modified force term is also proposed and included in the evolution equation of the DI-LBM. Some classical problems are used to test the DI-LBM, and the results are in good agreement with some available theoretical and numerical works. Finally, it is also found that the DI-LBM is more efficient and accurate than the PS-LBM with the superposition model.

\end{abstract}

\end{frontmatter}

\section{\label{sec:level1}Introduction}
Particulate flows are ubiquitous in both nature and engineering, such as, the sediment deposition, fluidized beds, and so on \cite{LI2011,Maxey2017}. With the rapid development of computer technology, the numerical simulation has been becoming an important and efficient tool in the study of the particulate flows. The lattice Boltzmann method (LBM), as a kinetic-based numerical approach, has gained a great success in the simulation of complex hydrodynamic problems \cite{Chen1998,Succi2001,Guo2013,Kruger2017}. Compared to the traditional methods for the Navier-Stokes equations, the LBM has some distinct advantages, including the clear physical background, easy implementation of boundary conditions, simplicity in programming and high computational efficiency \cite{Chen1998}. Due to these advantages of the LBM, it has also been applied to investigate particulate flows  \cite{Aidun,Ladd1994,Ladd2001}. Generally speaking, in the framework of LBM, there are three basic methods in the treatment of fluid-particle interface, namely, the bounce-back method, the immersed boundary (IB) method and the partially saturated (PS) method.

Ladd \cite{Ladd1994,Ladd2001} first proposed a shell method to simulate fluid-particle flows based on the half-way bounce-back scheme. In the bounce-back scheme, the inside and outside of the particle are filled with fluid, and the same bounce-back operation is performed for both the internal and external fluids of the particle. However, this method requires the particle boundary to be in the middle of the grid, which may cause the simulated particle boundary to be different from the actual physical boundary. To overcome this problem, several interpolation-based curved boundary schemes have been developed and applied to describe the particle-fluid interaction \cite{Yu2003,Bouzidi2003,zhao2017}. In this kind of method, there is no fluid inside the particle, and it has a second-order accuracy for arbitrary curved boundaries. However, the hydrodynamic force at the boundary is usually not smooth, and the fictitious oscillation is more serious when the moving particle is considered \cite{Tao2016}.

The IB method was first developed by Peskin \cite{Peskin1977} to study blood flows in the heart. Later, Feng \cite{Feng2004} proposed an immersed-boundary lattice Boltzmann method (IB-LBM) to simulate fluid-particle interaction problems. In this method, the Euler mesh is used for fluid domain, while the particle boundary is marked by a set of Lagrangian points. The fluid-particle interaction is achieved by adding an external force to the fluid with the Dirac function. However, the IB-LBM usually dose not satisfy the no-slip boundary condition \cite{Cai2018}. To enforce the no-slip boundary at the fluid-particle interface, some implicit IB-LBMs have also been developed \cite{Wu2009, Hu2014, Cai2018}.

The partially saturated (PS) method coupled with the LBM (hereafter the PS-LBM) was proposed by Noble and Torczynski \cite{Noble1998}, and then it is also applied to study some two-dimensional flows by Cook et al. \cite{Cook2004}. In this method, the fluid is filled with the whole domain. However, unlike above two methods, the fluid-particle interaction in this method is realized through modifying the collision term in the evolution equation of LBM where a parameter representing the volume fraction of solid phase is introduced. According to the modified collision term, there are two different models, i.e., the non-equilibrium bounce back (BB) and the superposition (SP) models. Compared to the bounce-back method \cite{Ladd1994,Ladd2001}, this method can give a relatively smooth hydrodynamic force.

Owing to the advantage in mass conservation and without special treatment on the fresh fluid nodes \cite{Kruger2017}, the PS-LBM has received increasing attention in studying the particulate flows \cite{Tsigginos2019,Najuch2019}. However, it should be noted that there is another problem in the PS-LBM, i.e., the accurate computation of the solid-phase volume fraction is very complicated, especially for three-dimensional problems. To eliminate this limitation of the PS-LBM, in this work a smooth function is first introduced to represent the volume fraction of solid phase, and then, based on the PS-LBM with SP model, we developed a diffuse-interface LBM (DI-LBM) for fluid-particle interaction problems. The rest of the paper is organized as follows. In Section~\ref{Numerical method}, a diffuse-interface lattice Boltzmann method is developed. In Section~\ref{Numerical results and discussion}, some numerical examples are carried out to test the present DI-LBM, and finally, some conclusions are given in Section~\ref{Conclusions}.

\section{Numerical method}\label{Numerical method}
In this section, we first presented a brief introduction to the standard LBM and PS-LBM, and then following the PS-LBM with SP model, a DI-LBM is developed for fluid-particle interaction problems.

\subsection{Standard lattice Boltzmann method}
In the standard LBM for fluid flows, the evolution equation reads \cite{Chen1998}
\begin{equation}
f_i(\mathbf{x}+\textbf{c}_i\Delta t,t+\Delta t)=f_i(\textbf{x},t)-{1 \over \tau}[f_i(\mathbf{x},t)-{f_i}^{eq}(\textbf{x},t)],
\label{eq:1}
\end{equation}
where $f_i(\textbf{x},t)$ is the distribution function at position $\textbf{x}$ and time $t$. $\{\mathbf{c}_i,i=0,\cdots,q-1\}$ is the set of discrete velocity, $q$ is the number of discrete velocity directions, $\Delta t$ is the time step, $\tau$ is the dimensionless relaxation time.
$f_i^{eq}(\textbf{x},t)$ is the equilibrium distribution function, and is defined by
\begin{equation}
f_i^{eq}(\textbf{x},t) = {\omega _i}\rho \left[ {1  + {{{\textbf{c}_i} \cdot \textbf{u}} \over {c_s^2}} + {({{\textbf{c}_i} \cdot \textbf{u}})^2 \over {2 c_s^4}}-{{{\textbf{u}^2} } \over {2c_s^2}}} \right],
\label{eq:2}
\end{equation}
where $\omega_i$ is the weight coefficient, ${c_s}$ is the lattice sound speed, and is related to the lattice speed $c$ ($c = {{\Delta x} \mathord{\left/{\vphantom {{\Delta x} {\Delta t}}} \right. \kern-\nulldelimiterspace} {\Delta t}}$, $\Delta x$ is the lattice spacing).

For simplicity but without loss of generality, in this work we only considered the two-dimensional problems, and adopted the commonly used D2Q9 lattice model \cite{Qian1992}, in which the discrete velocities are defined as
\begin{equation}
\textbf{c}_i=
\begin{cases}
(0,0), & i=0,\\
(\cos[(i-1)\pi /2],\sin[(i-1)\pi /2])c, & i=1,2,3,4,\\
\sqrt2(\cos[(2i-1)\pi /4],\sin[(2i-1)\pi /4])c, & i=5,6,7,8.
\end{cases}
\label{eq:23}
\end{equation}
The corresponding weight coefficients are given by
\begin{equation}
w_i=
\begin{cases}
4/9, & i=0,\\
1/9, & i=1,2,3,4,\\
1/36, & i=5,6,7,8.
\end{cases}
\label{eq:23}
\end{equation}

In addition, the macroscopic variable $\rho$ and $\textbf{u}$ can be computed by
\begin{equation}
\rho=\sum_i f_i, \;  \rho \textbf{u}=\sum_i \textbf{c}_i f_i.
\label{eq:3}
\end{equation}
Finally, through the Chapman-Enskog analysis \cite{Guo2013}, the Navier-Stokes equations can be recovered from Eq. (1) with the following viscosity,
\begin{equation}
\upsilon=c_s^2\left(\tau-{1\over2}\right)\Delta t.
\label{eq:4}
\end{equation}

\subsection{The partially saturated lattice Boltzmann method}
In the PS-LBM, the fluid-particle interaction is realized by introducing
an additional collision term $\Omega_i^s$, and consequently, the evolution equation can be written as
 \begin{equation}
 f_i(\mathbf{x}+\textbf{c}_i\Delta t,t+\Delta t)=f_i(\textbf{x},t)+(1-B)\left\{-{1\over \tau}[f_i(\mathbf{x},t)-{f_i}^{eq}(\textbf{x},t)]\right\}+B\Omega_i^s,
 \label{eq:5}
 \end{equation}
where $B$ is a weighting function of the solid-phase volume fraction $\varepsilon_s$, and usually it can be simply given by
 \begin{equation}
 B(\varepsilon_s,\tau)=\varepsilon_s.
 \label{eq:6}
 \end{equation}
 It can be find that $\varepsilon_s=0$ ($B=0$) denotes the pure fluid, and in this case, Eq. (7) would degenerate into the standard LBM (1); while $\varepsilon_s=1$ ($B=1$) represents the pure solid. For the SP model in the PS-LBM,  the additional collision term $\Omega_i^s$ in Eq. (7) is designed as
 \begin{equation}
 \Omega_i^s=f_{i}^{eq}(\rho,\textbf{u}_s)-f_i(\textbf{x},t)+\big(1-{1\over \tau}\big)[f_i(\textbf{x},t)-f_{i}^{eq}(\rho,\textbf{u})],
 \label{eq:7}
 \end{equation}
where $\textbf{u}_s$ is the velocity of the particle.

The total hydrodynamic force and torque imposed on solid particle are calculated by summing the additional collision operator over the nodes covered by the solid,
 \begin{equation}
 \textbf{F}_f=-{\Delta x^2 \over \Delta t} \sum_n B_n \sum_i \Omega_i^s \textbf{c}_i,
 \label{eq:8}
 \end{equation}
  \begin{equation}
  \textbf{T}_f=-{\Delta x^2 \over \Delta t} \sum_n (\textbf{x}_n-\textbf{x}_s) \times \left(B_n \sum_i \Omega_i^s \textbf{c}_i \right),
  \label{eq:9}
  \end{equation}
where $\textbf{x}_n$ is the coordinate of the node $n$, and $\textbf{x}_s$ is the position of the particle. In addition, we would also like to point out that Eq. (7) can also be written into another form \cite{Najuch2019},
  \begin{equation}
  f_i(\mathbf{x}+\textbf{c}_i\Delta t,t+\Delta t)=f_i(\textbf{x},t)-{1\over \tau}[f_i(\mathbf{x},t)-{f_i}^{eq}(\textbf{x},t)]+B[f_{i}^{eq}(\rho,\textbf{u}_s)-f_{i}^{eq}(\rho,\textbf{u})].
  \label{eq:5}
  \end{equation}
  It is clear that compared to the evolution equation (1) in the standard LBM, only an additional term $B[f_{i}^{eq}(\rho,\textbf{u}_s)-f_{i}^{eq}(\rho,\textbf{u})]$ is included in Eq. (12).

\subsection{A diffuse-interface lattice Boltzmann method}
From above discussion, one can find that the PS-LBM remains two advantages of the standard LBM, i.e., the locality of the collision operator and simple linear streaming operator, and thus it is easy to solve the problems involving the moving boundaries \cite{owen2011}. However, the accurate calculation of local solid-phase volume fraction $\varepsilon_s$ in the PS-LBM is complicated, and the computational cost is also expensive. To overcome these problems, here we introduce a parameter $\phi$, similar to the order parameter in the diffuse-interface method for multiphase flows \cite{shen2012}  or the solid-phase volume fraction $\varepsilon_s$ in the PS-LBM, and develop a new DI-LBM for particulate flows. Similar to the PS-LBM, in the DI-LBM, the sharp fluid-particle interface is replaced by a thin but nonzero transition region where the physical variables change continuously. In addition, based on the PS-LBM with the SP model \cite{Noble1998} or Eq. (12), the evolution of the DI-LBM is written as
 \begin{equation}
  f_i(\mathbf{x}+\textbf{c}_i\Delta t,t+\Delta t)=f_i(\textbf{x},t)+\left\{-{1\over \tau}[f_i(\mathbf{x},t)-{f_i}^{eq}(\textbf{x},t)]\right\}+\phi F_i,
  \label{eq:10}
  \end{equation}
where the parameter $\phi$ is a hyperbolic tangent function, and is defined as
 \begin{equation}
 \phi={1+\tanh(2l/\varepsilon)\over 2},
 \label{eq:11}
 \end{equation}
where $l$ is the distance to the boundary, $\varepsilon$ is the thickness of diffuse interface. $F_i$ is the discrete force term used to replace the last term $[f_{i}^{eq}(\rho,\textbf{u}_s)-f_{i}^{eq}(\rho,\textbf{u})]$ in Eq. (12), and is given by
 \begin{equation}
 F_i=\big(1-{1 \over 2\tau}\big)w_i\rho\left({\textbf{c}_i \cdot (\textbf{u}_s-\textbf{u}^*) \over c_s^2}+{(\textbf{u}_s\textbf{u}_s-\textbf{u}^*\textbf{u}^*): (\textbf{c}_i \textbf{c}_i-c_s^2 \textbf{I}) \over 2c_s^4}\right),
 \label{eq:12}
 \end{equation}
 where $\textbf{u}^{*}$ is the velocity without considering the fluid-particle interaction, $\textbf{u}$ is the corrected velocity, and they are defined by
  \begin{equation}
\textbf{u}^{*}={\sum_i \textbf{c}_i f_i \over \rho},
  \label{eq:12}
 \end{equation}
  \begin{equation}
\textbf{u}=\textbf{u}^{*}+{1 \over 2}\phi (\textbf{u}_s-\textbf{u}^{*}).
  \label{eq:12}
 \end{equation}

 The hydrodynamic force $\textbf{F}_f$ and torque $\textbf{T}_f$ can be calculated through the first-order moment of the discrete force term,
 \begin{equation}
 \textbf{F}_f=-{\Delta x^2 \over \Delta t} \sum_n \phi_n \sum_i F_i \textbf{c}_i,
 \label{eq:12}
  \end{equation}
 \begin{equation}
 \textbf{T}_f=-{\Delta x^2 \over \Delta t} \sum_n (\textbf{x}_n-\textbf{x}_s) \times \left(\phi_n \sum_i F_i \textbf{c}_i \right).
 \label{eq:13}
 \end{equation}

The calculation procedure of the DI-LBM is as follows:
\begin{enumerate}
    \item
     Initialize the values of variables.
    \item
     Compute the distribution functions in entire computational domain using Eq. (13).
     \item
     Correct the velocity field using Eq. (17).
     \item
     Calculate the hydrodynamic force and torque according to Eqs. (18) and (19).
     \item
     Update the velocity of the boundary and the parameter $\phi$.
     \item
     Repeat steps 2-5 until convergence is reached.
\end{enumerate}

 It should be noted that in the DI-LBM, a smooth function $\phi$ is used to replace the volume fraction of solid phase in the PS-LBM, and the difficulty in the computation of the solid-phase volume fraction can be avoided. Therefore, compared to the original PS-LBM with the SP model, the present DI-LBM would be more efficient. In addition, we proposed a modified discrete force term [see Eq. (15)], which also makes the present LB method more accurate than the original PS-LBM in the study of the particulate flows (see the numerical results in the following section). Through the Chapman-Enskog analysis (see the details in the Appendix), we can obtain the macroscopic equations of the present DI-LBM, which are similar to those of the IB-LBM \cite{Cai2018}.

\section{Numerical results and discussion}\label{Numerical results and discussion}
In this section, we conducted some numerical tests to validate the present DI-LBM; including the fluid flows passing a stationary circular cylinder, a particle settling along channel centerline, the off-centerline sedimentation of a particle, the sedimentation of two particles and a neutrally buoyant particle moving in the Poiseuille flow. In the following simulations, the non-equilibrium extrapolation scheme \cite{Guo2002} is applied to treat the physical boundary conditions of above problems.

\subsection{The fluid flows passing a stationary circular cylinder}
\begin{figure}[ph]
\centering \includegraphics[scale=0.75]{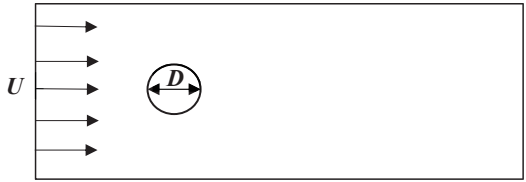}
\caption{The schematic of fluid flows passing a stationary circular cylinder.}
\label{Exp1_Fig1_0}
\end{figure}

We first considered the problem of fluid flows passing a stationary circular cylinder, and the schematic of the problem is shown in Fig. \ref{Exp1_Fig1_0}. It is known that the flow behavior of the problem is mainly governed by the Reynolds number ($Re$), which is a dimensionless number defined by
 \begin{equation}
 Re={U D \over \upsilon},
 \label{eq:16}
 \end{equation}
where $U$ is the free stream velocity, $D$ is the diameter of the cylinder, $\upsilon$ is the kinematic viscosity. In addition, for a specified case where the density $\rho$, the diameter $D$ and the velocity $U$ are given, we can calculate the drag coefficient,
 \begin{equation}
C_D={F_{fx} \over 0.5\rho D U^2},
 \label{eq:17}
 \end{equation}
 where $F_{fx}$ is the $x$-component of the hydrodynamic force.

\begin{figure}[ph]
 \subfigure[]{ \label{fig:mini:subfig:a}
 \includegraphics[scale=0.35]{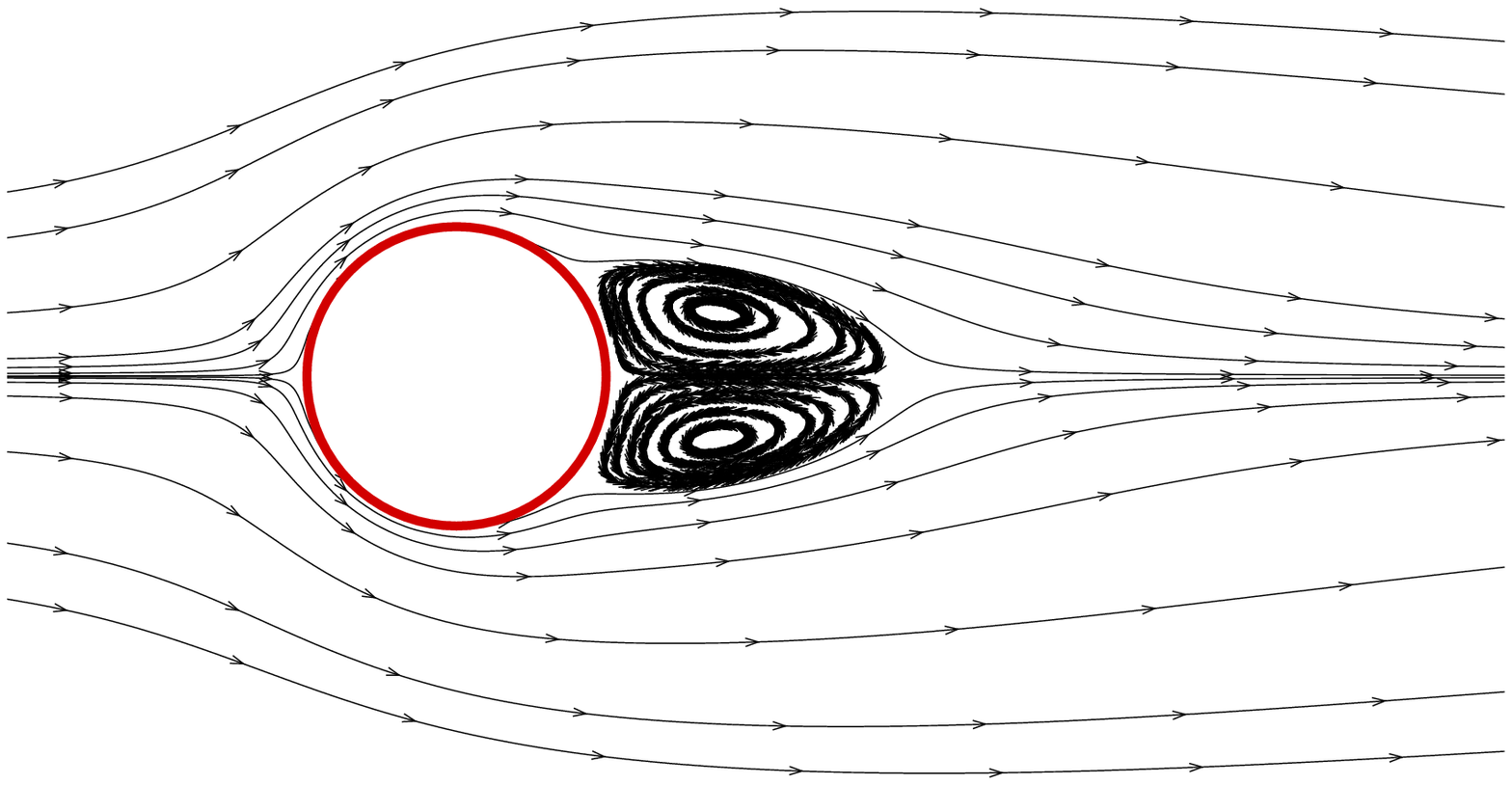}}
 \subfigure[]{ \label{fig:mini:subfig:b}
 \includegraphics[scale=0.35]{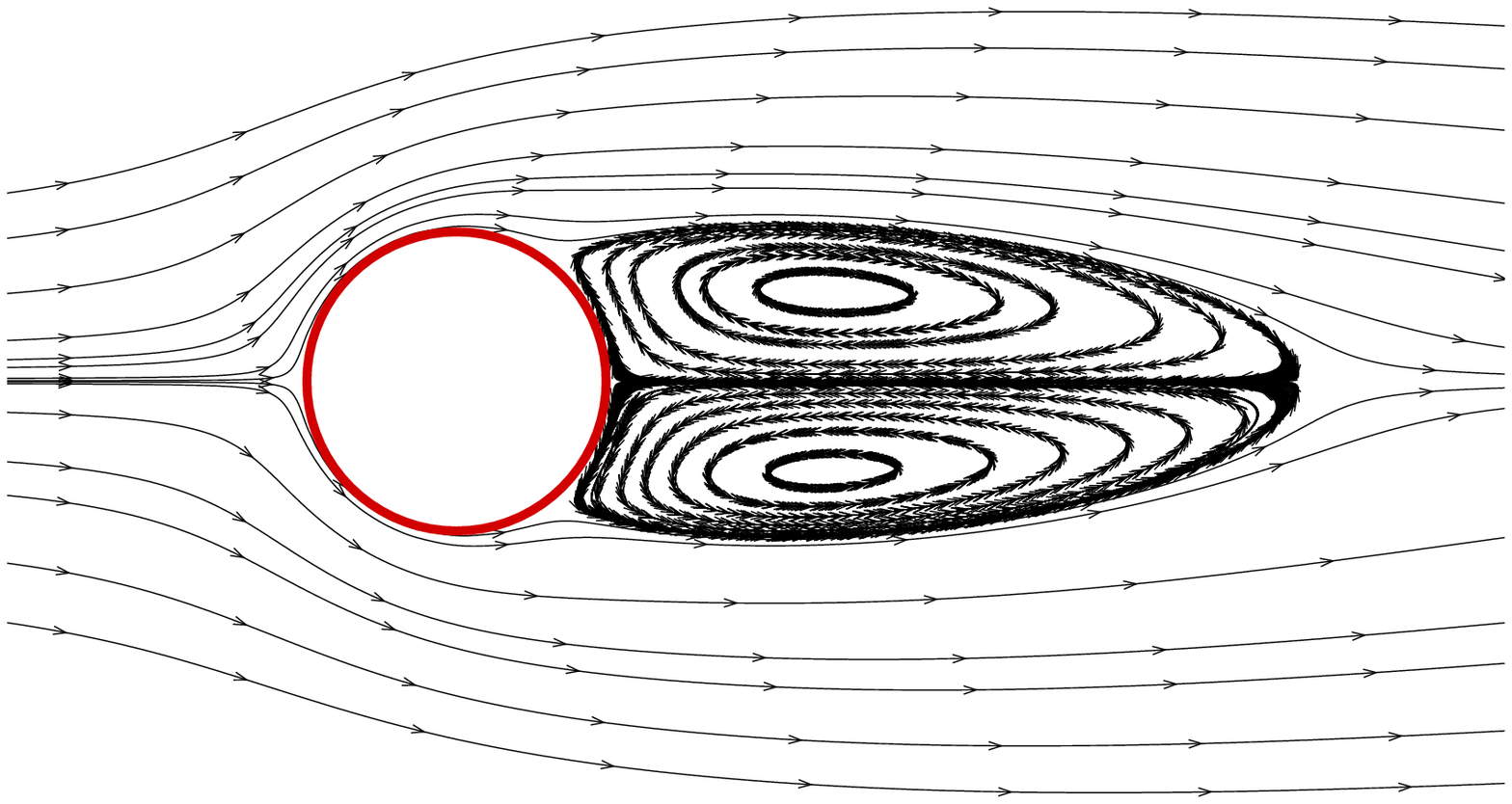}}
 \caption{The streamlines of the fluid flows around the stationary circular cylinder at $Re=20$ (a) and $40$ (b).}
 \label{Exp1_Fig1_1}
 \end{figure}

In our simulations, the computational domain is $35D \times 20D$, the circular cylinder with the diameter $D=1.0$ is placed at $(10D,\,10D)$, the free stream velocity is set to be $U=0.1$. We performed some simulations, and plotted the streamlines in Fig. \ref{Exp1_Fig1_1} where $Re=20$ and $40$. From this figure, it can be observed that when the flow field reaches steady state, there is a pair of symmetric recirculating eddies formed behind the cylinder, and the length of the recirculating region increases with the increase of $Re$. We noted that these results are qualitatively consistent with the previous work \cite{Niu2006}. In addition, to give a quantitative comparison between the present results and some available data \cite{Niu2006,He1997}, we also calculated the drag coefficient and dimensionless recirculation length $L_w=2L/D$ ($L$ is the recirculation length), and presented them in Table \ref{Tab_Exp1_1}. It can be seen from this table that the results of the present DI-LBM are in good agreement with those reported in the previous studies \cite{Niu2006,He1997}.

\begin{table}[ht]
\caption{A comparison of the drag coefficient and recirculation length between the present work and some previous studies.}
\centering
{\begin{tabular}{@{}lllllcc@{}}
\hline
$Re$   &&References   & $C_D$    & $2L/D$ \\ \hline
$Re=20$ &&Present      & $2.172$     & $1.875$   \\
        &&Niu et al. \cite{Niu2006}   & $2.144$    & $1.89$   \\
        &&He et al. \cite{He1997}   & $2.152$    & $1.842$  \\ \hline
$Re=40$ &&Present      & $1.64$     & $4.796$   \\
        &&Niu et al. \cite{Niu2006}  & $1.589$    & $4.52$   \\
        &&He et al. \cite{He1997}  & $1.499$    & $4.490$  \\ 
\hline
\end{tabular}
\label{Tab_Exp1_1}}
\end{table}

\subsection{A particle settling along channel centerline}
\begin{figure}[ph]
 \centering \includegraphics[scale=0.5]{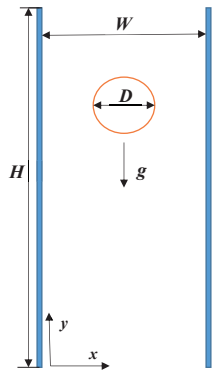}
 \caption{The configuration of a particle settling along channel centerline.}
 \label{Exp2_Fig2_0}
 \end{figure}

We continued to study the problem of a particle settling along channel centerline. The configuration of this problem is depicted in Fig. \ref{Exp2_Fig2_0} where a particle with the diameter $D$ is placed in the center of the channel, the width of the channel is $W$. Initially, the particle is released from rest, and then it would sink under the gravity force. It should be noted that unlike the previous problem, besides the Navier-Stokes equations, we also need to solve the following equations for particle motion,
\begin{equation}
m_s{d \textbf{u}_s \over dt}=\textbf{F}_f+\left({\rho_f \over \rho_s}-1\right)m_s\textbf{g},
\label{eq:30}
\end{equation}
\begin{equation}
I_s{d \omega_s \over dt}=T_f,
\label{eq:31}
\end{equation}
\begin{equation}
{d \textbf{x}_s \over dt}=\textbf{u}_s,
\label{eq:32}
\end{equation}
where $\textbf{u}_s$ is the velocity of the setting particle, $m_s$ and $\omega_s$ are the mass and the angular velocity of the particle, $\textbf{g}$ is the gravity acceleration, $I_s$ is the rotational inertia of the particle.

Theoretically, for the problem with a small Reynolds number, one can obtain the approximate solution of the steady drag force \cite{Happel1965},

 \begin{equation}
\textbf{F}_d=4\pi K \mu \textbf{u}_s.
 \label{eq:17}
 \end{equation}
The parameter $K$ is a correction factor, and is given by
 \begin{equation}
K=(\ln W^*-0.9175+1.7244W^{* -2}-1.7302W^{* -4}+2.4056W^{*-6}-4.5913W^{*-8})^{-1},
 \label{eq:17}
 \end{equation}
where $W^*=W/D$ is the blockage ratio.

When the problem reaches the equilibrium state, we can also derive the final velocity of the settling particle \cite{wang2013},
 \begin{equation}
 \textbf{u}_s={D^2 \over 16 K \mu}(\rho_f- \rho_s )\textbf{g}.
 \label{eq:17}
 \end{equation}

In our simulations, the diameter of the cylinder $D$ is set to be $0.24\,\rm cm$, the computational domain is $W \times H=1.2\, \rm cm \times 6.0\, \rm cm$, the density of the fluid is $\rho_f=1.0\, \rm g/\rm cm^3$, and the viscosity of the fluid is $\mu=0.1\, \rm g/(\rm cm \cdot \rm s)$. At the initial time $t=0$, the particle is released at the point $(0.6\, \rm cm,\,3.0\,\rm cm)$. For the left and right channel walls, the no-slip boundary condition is applied. At the inlet, the velocity is set to be zero, while at the outlet, the fully developed boundary conditions is adopted.

 \begin{figure}[ph]
 \centering \includegraphics[scale=0.5]{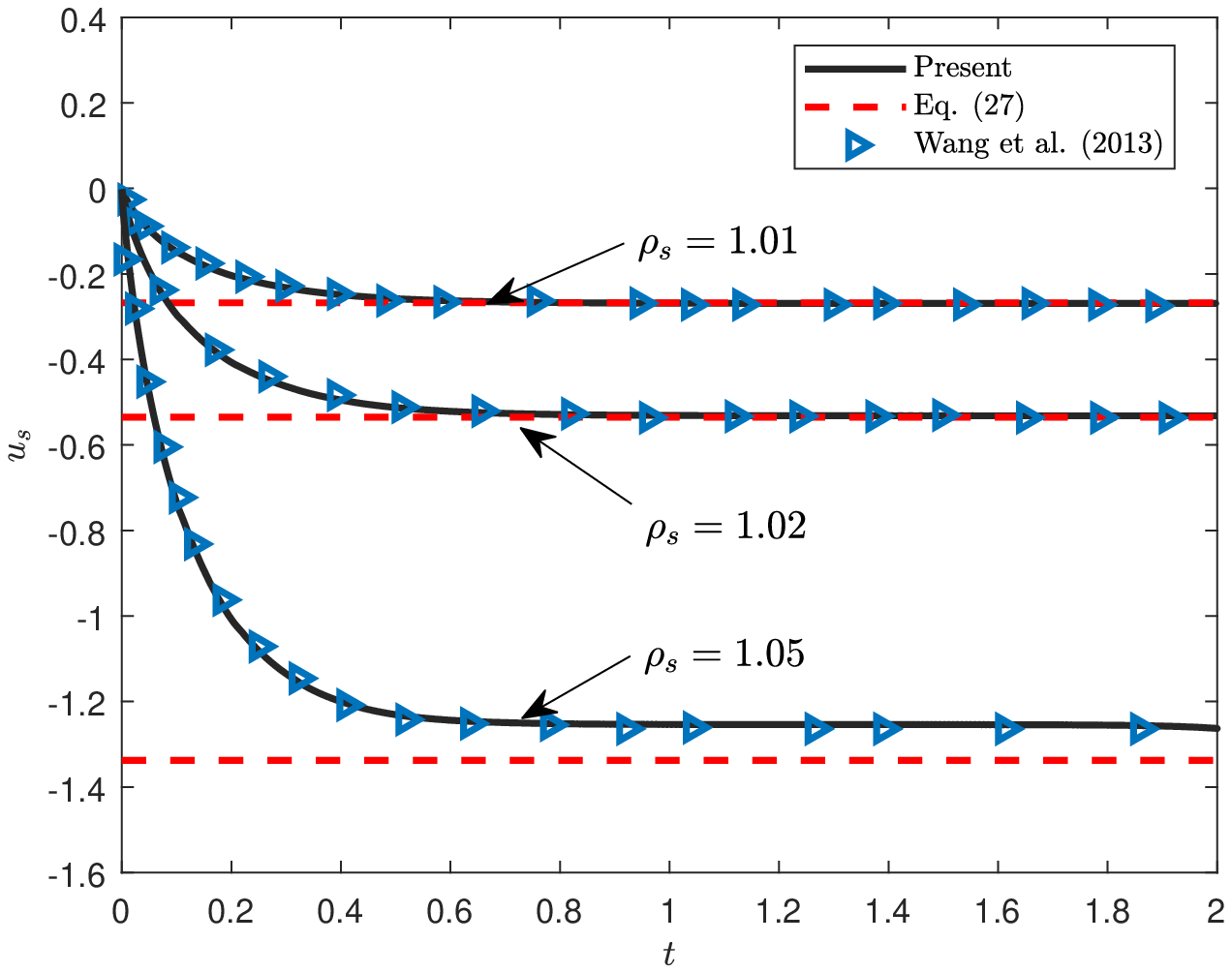}
 \caption{The settling velocities of the particle at different values of the particle density ($\rho_s =1.01,1.02$ and $1.05 \rm g/\rm cm^3$).}
 \label{Exp2_Fig2_1}
 \end{figure}

 We carried out some simulations with three different values of the particle density ($\rho_s=1.01,1.02$ and $1.05\,\rm g/\rm cm^3$), and presented the settling velocities of three different cases in Fig. \ref{Exp2_Fig2_1} where the grid size is $\Delta x=0.01$ and the relaxation time is $\tau=0.8$. As shown in this figure, the velocity of the settling particle increases with the increase of particle density, and the present results also agree well with the previous numerical data \cite{wang2013}. In addition, it is also found that when the particle density is small, the numerical results are very close to the theoretical solution (27), while if the particle density becomes large (e.g., $\rho_s=1.05\,\rm g/\rm cm^3$), there would be a deviation between the numerical and theoretical results. This is because with the increase of the particle density, the settling velocity of the particle and the Reynolds number would be increased, which may cause the theoretical prediction (27) to be inaccurate.

\subsection{The off-centerline sedimentation of a particle}
\begin{figure}[ph]
 \centering \includegraphics[scale=0.5]{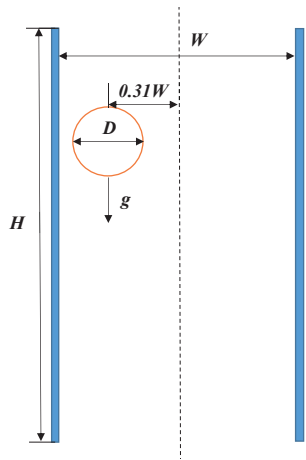}
 \caption{The schematic of the off-centerline sedimentation of a particle settling in a channel.}
 \label{Exp3_Fig3_0}
 \end{figure}

Next, we considered the sedimentation of a particle off the centerline. The schematic of the problem is shown in Fig. \ref{Exp3_Fig3_0} where a particle with the diameter $D=0.1\,\rm cm$ is located at the off-center line of the channel. The width and height of the channel are $W=4D$ and $H=400/13D$, the initial position of the particle is $(0.19W,\, 0.75H)$, the density of the fluid is $\rho_f=1.0\,\rm g/\rm cm^3$, and the viscosity of fluid is $0.01\,\rm cm^2/s$. Similar to above discussion, here two different values of the particle density, i.e., $\rho_s=1.01$ and $1.03\,\rm g/\rm cm^3$ are considered.

 \begin{figure}[h]
 \subfigure[]{ \label{fig:mini:subfig:a}
 \includegraphics[scale=0.5]{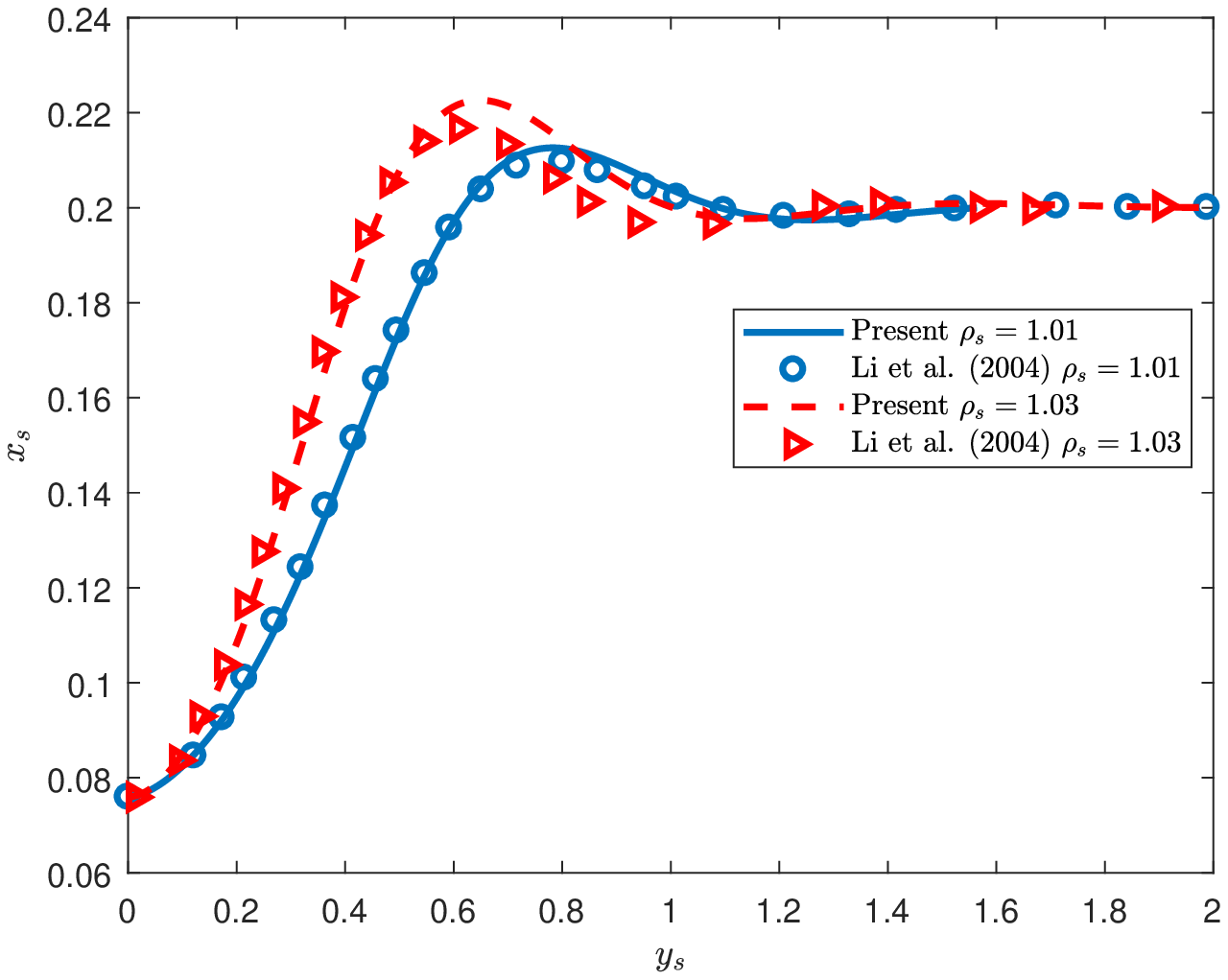}}
 \subfigure[]{ \label{fig:mini:subfig:b}
 \includegraphics[scale=0.5]{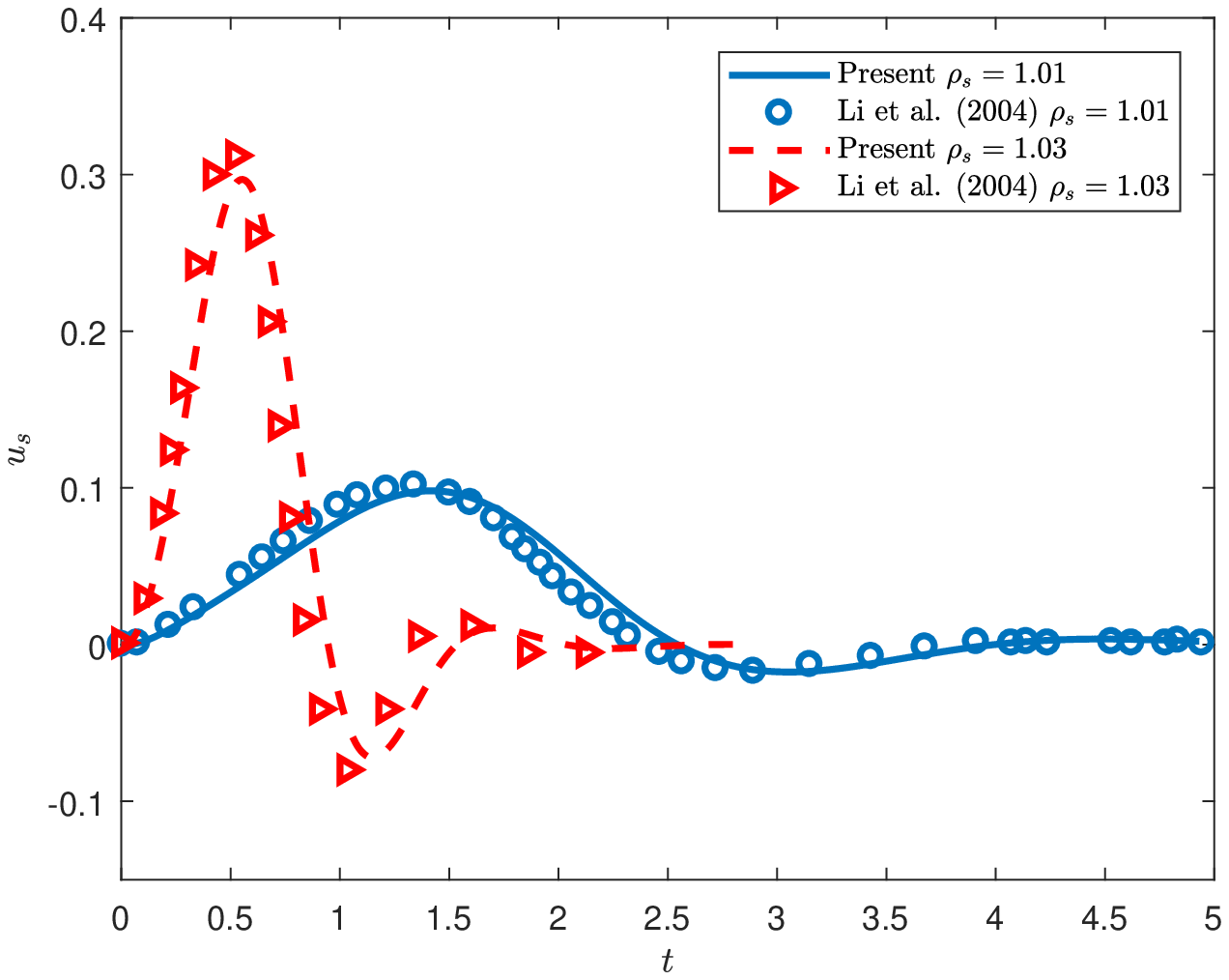}}
  \subfigure[]{ \label{fig:mini:subfig:c}
  \includegraphics[scale=0.5]{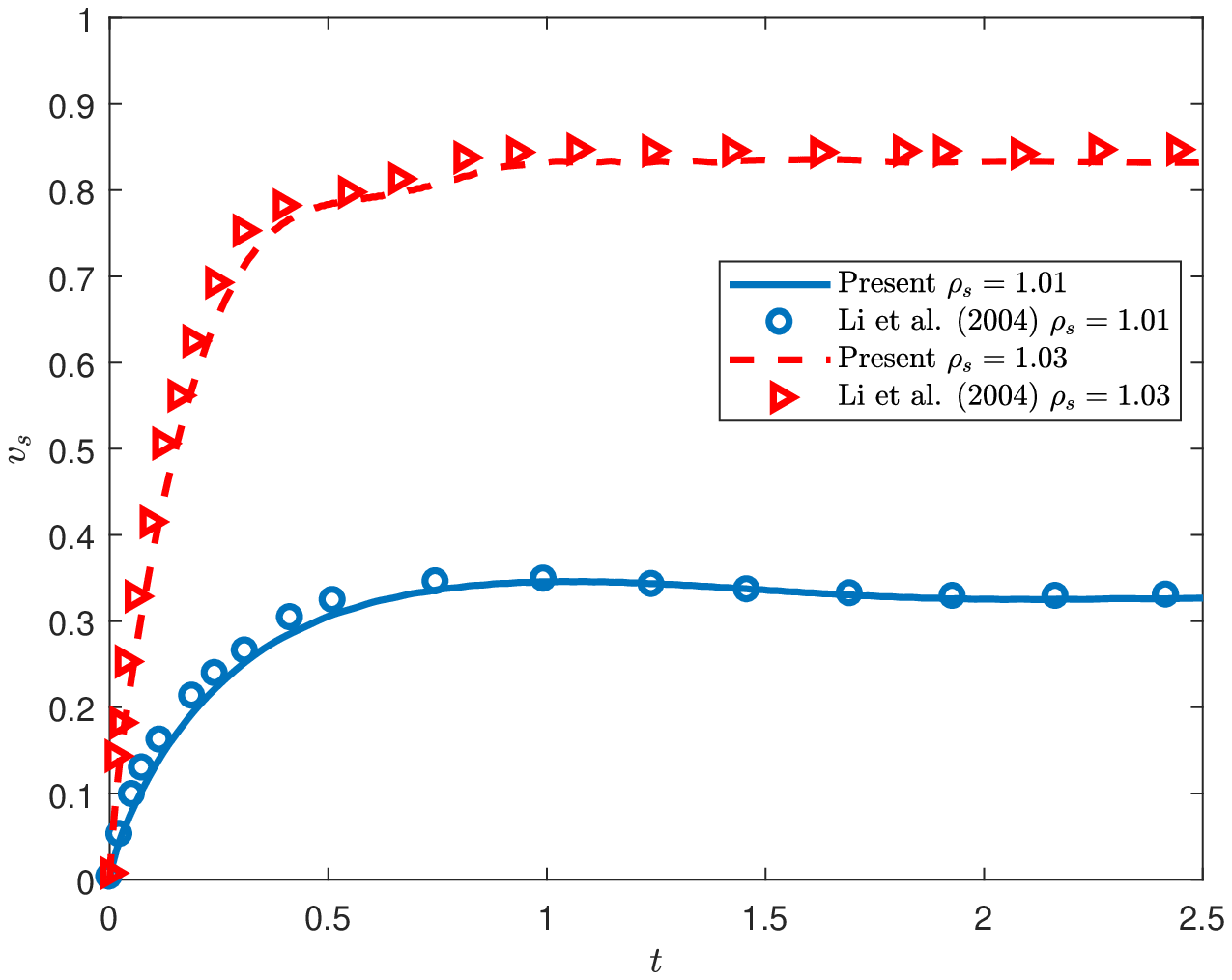}}
  \subfigure[]{ \label{fig:mini:subfig:d}
  \includegraphics[scale=0.5]{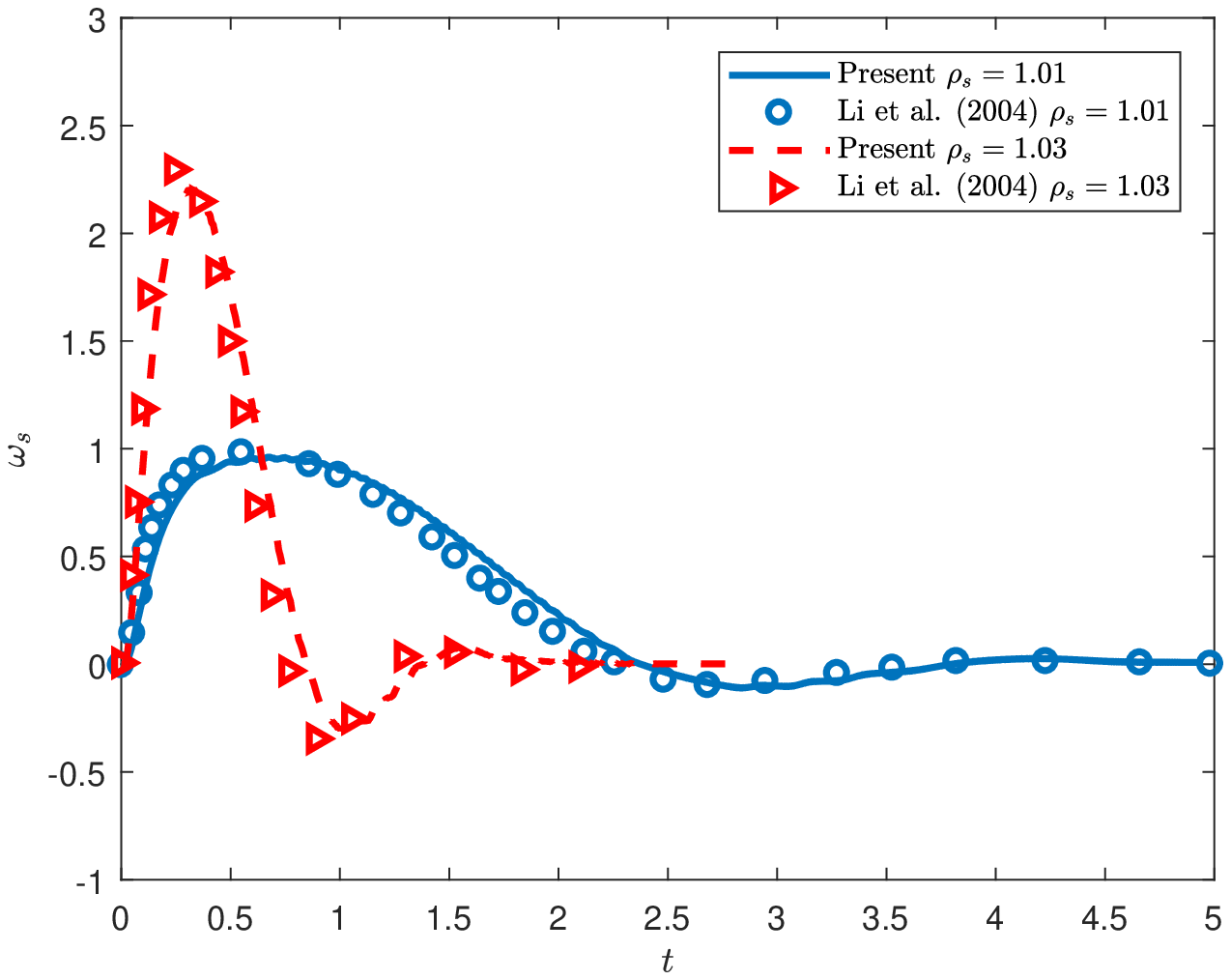}}
 \caption{ The trajectory of the particle and time history of the particle velocity [(a): trajectory; (b): $x$-component velocity; (c): $y$-component velocity; (d): angular velocity].}
 \label{Exp3_Fig3_1}
 \end{figure}

We conducted some simulations with the lattice size $Nx\times Ny=104\times 800$ and the relaxation time $\tau=0.6$, and presented the trajectory of the particle denoted by $(x_{s}, y_{s})$ and time history of the particle velocity in Fig. \ref{Exp3_Fig3_1}. Based on the results shown in Fig. \ref{Exp3_Fig3_1}(a), one can find that the particle finally moves to the centerline of the channel $(x_c=0.2\, \rm cm)$, which is due to the asymmetric force induced by the channel walls. From Fig. \ref{Exp3_Fig3_1}(b) and \ref{Exp3_Fig3_1}(d), we can also observe that with the increase of time, the horizontal velocity $u_s$ and the angular velocity $\omega_s$ gradually become zero. This is because when the time is large enough, the particle would reach the steady state along the channel centerline. In addition, the results in Fig. 6(c) indicate that with the increase of the particle density, the vertical velocity $v_{s}$ increase owing to the large gravity. Finally, we would like to point out that the present results agree well with the numerical data reported in the previous work \cite{Li2004}.

\subsection{The sedimentation of two particles}
To further test the capacity of the DI-LBM in the study of the particulate flows, the problem of the sedimentation of two particles is also considered. For this problem, the diameter of the particles is $D=0.2\, \rm cm$, the channel size is $2 \, \rm cm \times 8 \, \rm cm$, the viscosity of fluid is $0.01\, \rm g/(\rm cm\cdot \rm s)$, the densities of the fluid and particles are $\rho_f=1.0\, \rm g/\rm cm^3$ and $\rho_s=1.01\, \rm g/\rm cm^3$. The initial positions of two particles are $(0.999\, \rm cm, \,7.2\, \rm  cm)$ and $(1.0\, \rm cm, \, 6.8\, \rm cm)$.

In the following simulations, to avoid the overlap between two particles or between the particles and the walls, some short-range repulsive forces should be included. In this work, we adopted the collision model proposed by Wan and Turek \cite{Wan2006}, in which the repulsive forces can be expressed as

\begin{equation}
F_{i,j}^{P-P}=
\begin{cases}
0, & d_{i,j}>r_i+r_j+\xi,\\
{1 \over \epsilon_p^{'}}(X_i-X_j)(r_i+r_j-d_{i,j}), & d_{i,j}\leq r_i+r_j,\\
{1 \over \epsilon_p}(X_i-X_j)(r_i+r_j+\xi-d_{i,j})^2, & r_i+r_j \leq d_{i,j}\leq r_i+r_j+\xi,
\end{cases}
\label{eq:23}
\end{equation}

\begin{equation}
F_{i}^{P-W}=
\begin{cases}
0, & d_{i}^{'}>2r_i+\xi,\\
{1 \over \epsilon_W^{'}}(X_i-X_i^{'})(2r_i-d_{i}^{'}), & d_{i}^{'}\leq 2r_i,\\
{1 \over \epsilon_W}(X_i-X_i^{'})(2r_i+\xi-d_{i}^{'})^2, & 2r_i \leq d_{i}^{'}\leq 2r_i+\xi,
\end{cases}
\label{eq:23}
\end{equation}
where $r_i$ is the radius of the $i$th particle, $d_{i,j}=| X_i-X_j|$, $d_{i}^{'}=| X_i-X_i^{'}|$. $X_i$ is the $i$th particle center, $X_i^{'}$ is the coordinate vector of the center of the nearest imaginary particle on the boundary wall. $\epsilon_P$ and $\epsilon_P^{'}$ are small positive stiffness parameters for particle-particle collisions, and are fixed as $1.0\times 10^{-7}$. $\epsilon_W$ and $\epsilon_W^{'}$ are two small positive stiffness parameters for particle-wall collisions, and are taken as $\epsilon_W=\epsilon_P/2$ and $\epsilon_W^{'}=\epsilon_P^{'}/2$. $\xi$ is the threshold that is set to be one lattice unit.

 \begin{figure}[ph]
 \centering \includegraphics[scale=0.5]{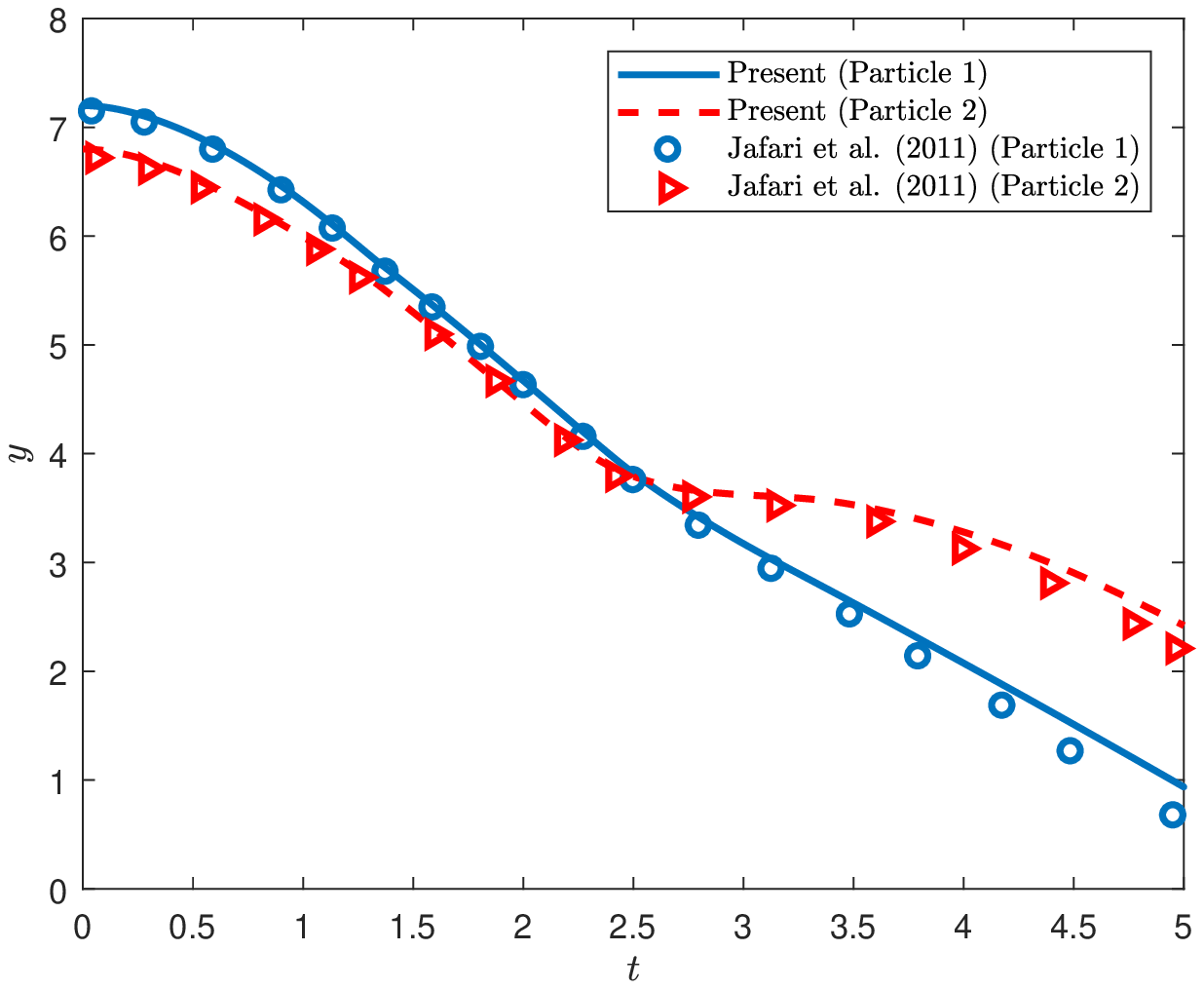}
 \caption{The vertical positions of the two particles.}
 \label{Exp4_Fig4_1}
 \end{figure}

 \begin{figure}[ph]
 \centering \includegraphics[scale=0.5]{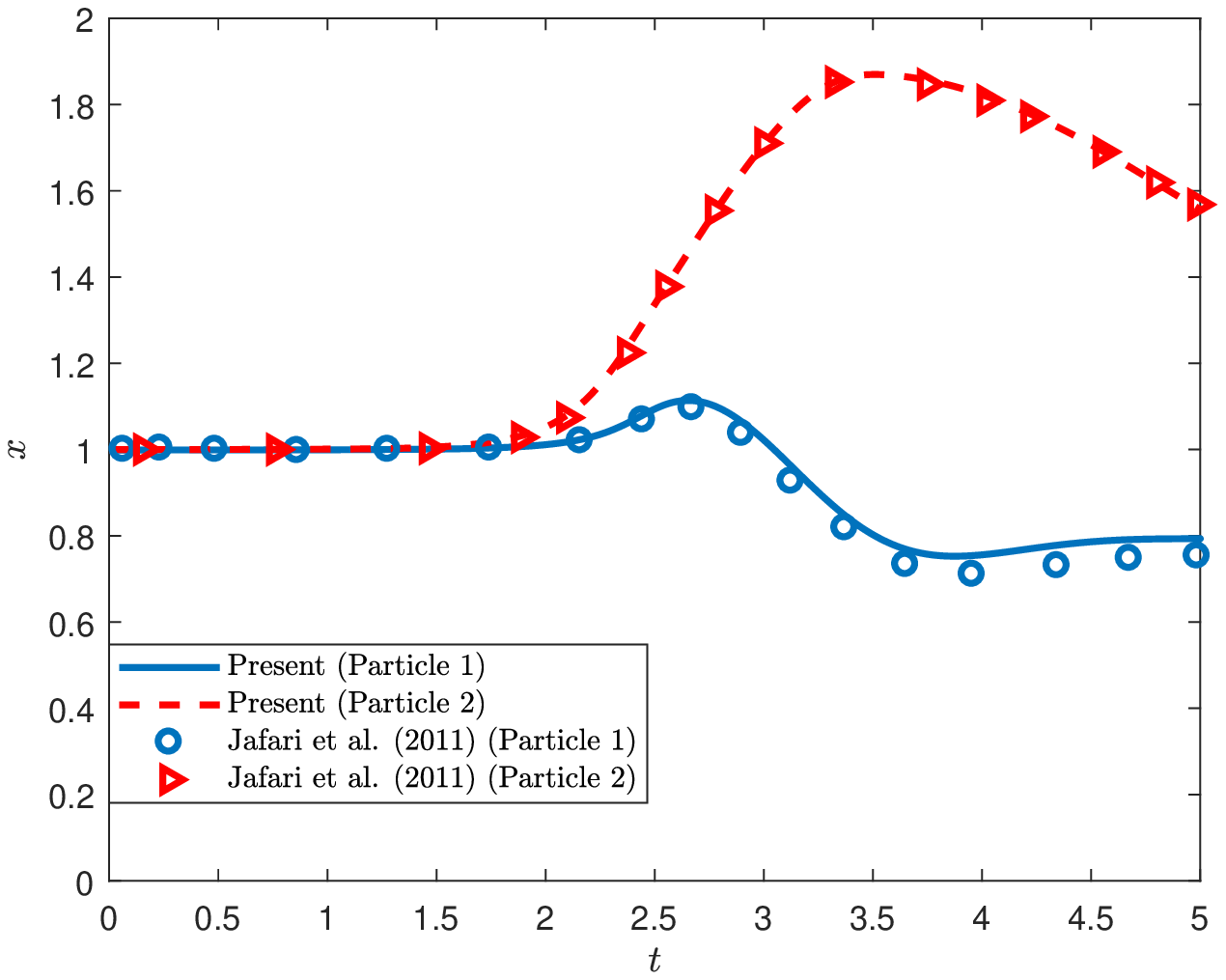}
 \caption{The horizontal positions of the two particles.}
 \label{Exp4_Fig4_2}
 \end{figure}

Figs. \ref{Exp4_Fig4_1} and \ref{Exp4_Fig4_2} show the vertical and horizontal positions of the two particles, which are also compared with the results of Jafari et al. \cite{Jafari2011}. From these two figures, one can find that as time goes on, the trailing particle first comes close to the leading particle because of the low pressure wake created by the leading particle, then the trailing particle with a larger velocity induces a kissing contact with the leading particle, and finally two particles tumble and separate from each other. These three distinct processes are so-called drafting, kissing and tumbling. In addition, it is also observed that the present results are close to those in the previous study \cite{Jafari2011}.

\subsection{A neutrally buoyant particle moving in the Poiseuille flow}
\begin{figure}[ph]
 \centering \includegraphics[scale=0.5]{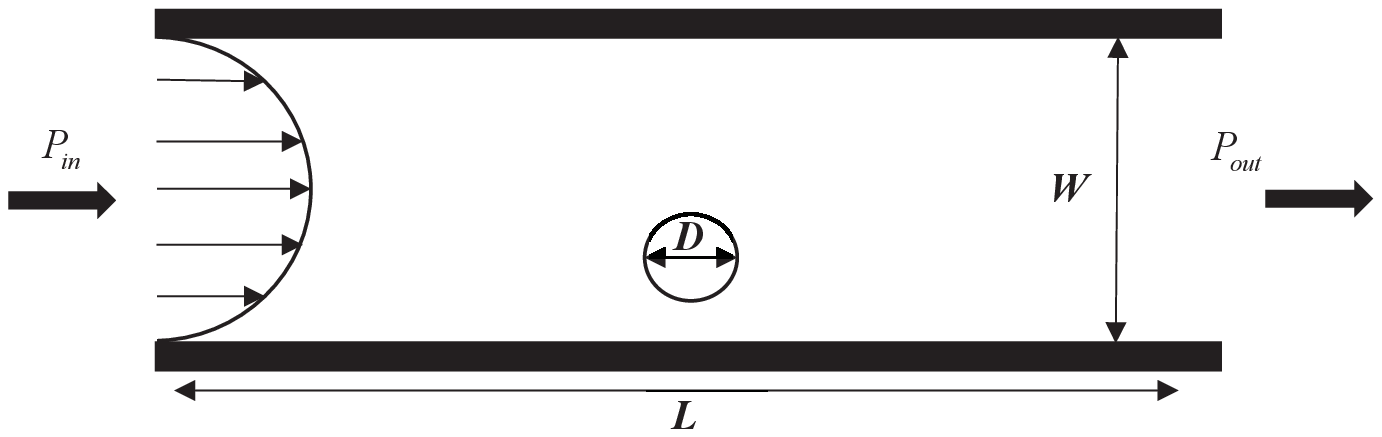}
 \caption{Schematic of a particle moving in the Poiseuille flow.}
 \label{Exp5_Fig5_0}
 \end{figure}

The last problem we considered is the motion of a neutrally buoyant particle in the Poiseuille flow. The configuration of this problem is depicted in Fig. \ref{Exp5_Fig5_0} where computational domain is $L \times W=20\times 4$, the diameter of the particle is $D=1$, and the initial position of the particle is $(x_0,y_0)=(L/2,\,0.25W)$. In order to reduce the computational cost, the moving computational domain is adopted, when the particle moves one lattice unit in the horizontal direction, the computational domain also moves one lattice in the same direction \cite{wang2013}. The pressure drop from inlet to outlet is $\triangle p=P_{in}-P_{out}=0.00267$, the lattice spacing is $\Delta x=1/25$, and the relaxation time is $\tau=0.75$.

 \begin{figure}[ph]
 \centering
\includegraphics[scale=0.5]{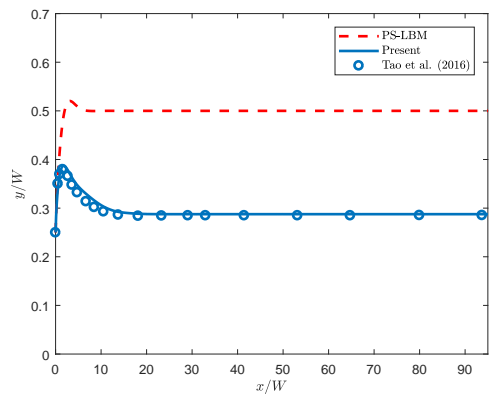}
\caption{The trajectory of the moving particle in the Poiseuille flow.} \label{Exp5_Fig1}
\end{figure}

\begin{table}[tbp]
\caption{A comparison of the equilibrium position of the particle between the present work and some previous studies. }
 \label{Tab_Exp5_1}
  \centering
\begin{tabular}{cccccccl}
\hline
Present &&  PS-LBM   &&  Li et al. \cite{Li2004} && Tao et al. \cite{Tao2016}\\
\hline

$0.2876$  && $0.50$  &&   $0.2874$  && $0.2849$\\

\hline
\end{tabular}
\end{table}

We carried out some simulations, and presented the trajectory of the particle in Fig. \ref{Exp5_Fig1}. As shown in this figure, with the increase of time, the particle would finally reach the equilibrium position between the wall and the centerline of the channel, which is also in good agreement with the previous work \cite{Tao2016}. However, when the original PS-LBM with the SP model is adopted to investigate this problem, the particle would reach the centerline of the channel, which may be caused by the inappropriate collision term or the discrete force term \cite{Noble1998,Najuch2019}. In addition, we also measured the equilibrium positions of the particle with the DI-LBM and PS-LBM with the SP model, and conducted a comparison of the values obtained with different numerical methods in Table \ref{Tab_Exp5_1}. From this table, one can also observe that the present result agrees well with the previous works \cite{Li2004,Tao2016}, while the original PS-LBM with the SP model cannot give the accurate result.

\section{Conclusions}\label{Conclusions}
In this work, we proposed a diffuse-interface lattice Boltzmann method for fluid-particle interaction problems. The distinct feature of the method is that the sharp boundary between the fluid and solid is represented by a diffuse interface with a finite thickness, and simultaneously a smooth function $\phi$ is introduced to describe the diffuse interface. This method can not only overcome the disadvantage of the PS-LBM in the computation of solid-phase volume fraction, but also improve the accuracy of the PS-LBM through developing a modified force term. The present DI-LBM is validated by several numerical experiments, and the results show that the DI-LBM is effective and accurate in the study of the fluid-particle interaction problems.

In addition, we would also like to point out that the present DI-LBM can be considered as an effective tool for more complex fluid-particle interaction problems, for instance, the moving particles in porous media \cite{su2019}.

\section*{Acknowledgements}
This work was financially supported by the National Natural Science Foundation of China (Grants No. 12072127, No. 11702259 and No. 51836003).

\section*{Appendix}\label{Appendix A}
\appendix
\setcounter{equation}{0}
\renewcommand{\theequation}{A.\arabic{equation}}
We now perform a detailed Chapman-Enskog analysis to obtain the macroscopic equations from the DI-LBM (13). In the Chapman-Enskog analysis, the distribution function, the time and space derivatives, and the force term can be first expanded as
\begin{subequations}
\label{eq:a1}
\begin{align}
f_i=f_i^{(0)}+\epsilon f_i^{(1)}+\epsilon ^2 f_i^{(2)}+\cdots,\\
\partial_t=\epsilon \partial_{t_1}+\epsilon^2 \partial_{t_2},\;\;\nabla=\epsilon \nabla_1,\;\;F_i=\epsilon F_i^{(1)},
\end{align}
\end{subequations}
where $\epsilon $ is a small parameter proportional to the Knudsen number.
Then, taking the Taylor expansion to Eq. (13) , we have
\begin{equation}
\Delta t D_i f_i+{\Delta t^2\over 2}{D_i}^2 f_i+\cdots=\left\{-{1\over \tau}[f_i(\mathbf{x},t)-{f_i}^{eq}(\textbf{x},t)]\right\}+\phi F_i.
\label{eq:a2}
\end{equation}
If we substitute Eq.~(\ref{eq:a1}) into Eq.~(\ref{eq:a2}), one can obtain
\begin{equation}
O(\epsilon^0):f_i^{(0)}=f_i^{eq}(\rho,\textbf{u}),
\label{eq:a3}
\end{equation}
\begin{equation}
O(\epsilon^1):\quad D_{1i}f_i^{(0)}=-{1 \over \tau \Delta t} f_i^{(1)} + {\phi \over \Delta t} F_i^{(1)},
\label{eq:a4}
\end{equation}
\begin{equation}
O(\epsilon^2):\quad \partial_{t_2}f_i^{(0)}+D_{1i}f_i^{(1)}+{\Delta t \over 2}D_{1i}^2f_i^{(0)}=-{1 \over \tau \Delta t} f_i^{(2)} ,
\label{eq:a5}
\end{equation}
where $D_{1i}=\partial_{t_1}+\mathbf{c}_i \cdot \nabla_1$.
According to Eq.~(\ref{eq:a3}), we can derive the following moments,
\begin{equation}
\begin{split}
&\sum_i f_i^{(0)}=\rho \\
&\sum_i \mathbf{c}_i f_i^{(0)}=\rho \textbf{u},\\%
&\sum_i \mathbf{c}_i \mathbf{c}_i f_i^{(0)}=c_s^2\rho \mathbf{I} + \rho \textbf{u}\textbf{u},\\
&\sum_i \mathbf{c}_i \mathbf{c}_i \mathbf{c}_i f_i^{(0)}=c_s^2 \Delta \cdot \rho \textbf{u} ,
\end{split}
\label{eq:a7}
\end{equation}
which can be used to obtain the zero and first-order moments of the non-equilibrium distribution functions,
\begin{equation}
\sum_i f_i^{(k)}=0\;(k\geq 1),\;\;
\sum_i \mathbf{c}_i f_i^{(1)}=-{\phi \over 2}\rho (\textbf{u}_s-\textbf{u}^*),\sum_i \mathbf{c}_i f_i^{(2)}=0.
\label{eq:a8}
\end{equation}
Substituting Eq.~(\ref{eq:a4}) into Eq.~(\ref{eq:a5}) yields
\begin{equation}
O(\epsilon^2):\quad \partial_{t_2}f_i^{(0)}+D_{1i}(1-{1 \over 2\tau})f_i^{(1)}=-{1 \over \tau \Delta t} f_i^{(2)} - D_{1i}{\phi \over 2}F_i^{(1)}.\label{eq:a9}
\end{equation}
Summing Eqs.~(\ref{eq:a4}) and ~(\ref{eq:a9}) over $i$, we have
\begin{equation}
\partial_{t_1}\rho+\nabla_1 \cdot (\rho \textbf{u})=0,
\label{eq:a10}
\end{equation}
\begin{equation}
\partial_{t_2}\rho=0.
\label{eq:a11}
\end{equation}
Combining Eqs.~(\ref{eq:a10}) and ~(\ref{eq:a11}), one can obtain the continuity equation,
\begin{equation}
\partial_{t}\rho+\nabla \cdot (\rho \textbf{u})=0.
\label{eq:a12}
\end{equation}
If we multiply $\mathbf{c}_i$ on both sides of Eqs.~(\ref{eq:a4}) and ~(\ref{eq:a9}), one can derive
\begin{equation}
\partial_{t_1}(\rho \textbf{u})+\nabla_1 \cdot (c_s^2\rho \mathbf{I} + \rho \textbf{u}\textbf{u})=\bar{F}^{(1)},
\label{eq:a13}
\end{equation}
\begin{equation}
\partial_{t_2}(\rho \textbf{u})+\nabla_1 \cdot \left(1-{1 \over 2\tau}\right) \sum \mathbf{c}_i \mathbf{c}_i f_i^{(1)}=- \nabla_1 \cdot {\phi \over 2} \sum \mathbf{c}_i \mathbf{c}_i F_i^{(1)},
\label{eq:a14}
\end{equation}
where $\bar{F}^{(1)}={\phi \over \Delta t}\rho(\textbf{u}_s-\textbf{u}^*)$.

With the help of Eq.~(\ref{eq:a4}), we can give an evaluation to second-order moment of $f_i^{(1)}$,
\begin{equation}
\begin{split}
-{1\over \tau \Delta t}\sum \mathbf{c}_i \mathbf{c}_i f_i^{(1)}
&=\sum \mathbf{c}_i \mathbf{c}_i D_{1i} f_i^{eq}-{\phi \over \Delta t} \sum \mathbf{c}_i \mathbf{c}_i F_i^{(1)} \\
&\approx c_s^2 \rho [\nabla_{1} \textbf{u}+(\nabla_{1} \textbf{u})^T].\\
\end{split}
\label{eq:a15}
\end{equation}
Substituting Eq.~(\ref{eq:a15}) into Eq.~(\ref{eq:a14}), and combining the equations at $t_1$ and $t_2$ scales, we have
\begin{equation}
\begin{split}
\partial_{t}(\rho \textbf{u})+\nabla \cdot (\rho \textbf{u}\textbf{u})
&=-\nabla P+ \nabla \cdot \left\{ \big(\tau-{1 \over 2}\big)\Delta t c_s^2\rho\left[\nabla \textbf{u}+(\nabla \textbf{u})^T \right]\right\}+\textbf{f},
\end{split}
\label{eq:a16}
\end{equation}
where $\textbf{f}$ is the force caused by the fluid-particle interaction.

From above discussion, one can find that the macroscopic equations ~(\ref{eq:a12}) and ~(\ref{eq:a16}) can be obtained from the DI-LBM with the following viscosity,
\begin{equation}
\mu=c_s^2\rho \left(\tau-{1\over2}\right)\Delta t.
\label{eq:a19}
\end{equation}

\section*{Acknowledgements}

\end{document}